\documentclass[aps,prd,preprint,preprintnumbers,nofootinbib,showpacs,floatfix,amsmath,amssymb]{revtex4}
\usepackage{mathrsfs}
\usepackage{amssymb}
\usepackage{amsmath}
\usepackage[amsmath,thmmarks]{ntheorem}

\usepackage{graphicx}
\usepackage{slashed}

\newcommand{\be}{\begin{eqnarray}}
\newcommand{\ee}{\end{eqnarray}}

\begin{document}

\title{Parton distributions and the $W$ mass measurement}

\author{Seth~Quackenbush}
\email{squackenbush@hep.fsu.edu}
\affiliation{Physics Department, Florida State University, Tallahassee, Florida
32306-4350, USA}

\author{Zack~Sullivan}
\email{Zack.Sullivan@IIT.edu}
\affiliation{Department of Physics, Illinois Institute of Technology, Chicago, Illinois 60616-3793, USA}

\preprint{IIT-CAPP-15-01}

\begin{abstract}
  We examine the sources of parton distribution errors in the $W$ mass
  measurement, and point out shortcomings in the existing
  literature. Optimistic assumptions about strategies to reduce the
  error by normalizing to $Z$ observables are examined and found to
  rely too heavily on assumptions about the parametrization and
  degrees of freedom of the parton distribution functions (PDFs). We
  devise a strategy to combine measurements as efficiently as possible
  using error correlations to reduce the overall uncertainty of the
  measurement, including $Z$ data, and estimate a PDF error of
  $^{+10}_{-12}$ MeV is achievable in a $W$ mass measurement at the
  LHC.  Further reductions of the $W$ mass uncertainty will require
  improved fits to the parton distribution functions.
\end{abstract}

\date{February 11, 2015}


\maketitle

\section{Introduction}
\label{intro}

As the Large Hadron Collider (LHC) at CERN begins its next run, there
is significant interest in reducing the uncertainty in the measured
value of the $W$ boson mass \cite{Baak:2013fwa}.  The current best
measurement of the $W$ boson mass of $80.385\pm 15$ GeV is based on
combined data from the CDF and D0 Collaborations
\cite{Aaltonen:2013iut} taken at the Fermilab Tevatron.  The
uncertainty in the experimental analyses is dominated by a theoretical
uncertainty generically called ``parton distribution function (PDF)
errors.''  Looking forward to a measurement at the LHC, we clarify how
``PDF errors'' affect the measurements, assess their contribution to
the uncertainty of the $W$ boson mass using current PDF sets, and
propose a series of steps to reduce that uncertainty by at least a
factor of three at the LHC.

Since the discovery of the $W$ boson in 1983 by the UA1 Collaboration
\cite{Arnison:1983rp}, the $W$ boson mass has played a central role in
precision electroweak measurements and in constraints on the standard
model through global fits.  For many years the uncertainty in the
measurement of the $W$ boson mass was one of the main limits to the
indirect prediction of the standard model Higgs boson mass
\cite{Agashe:2014kda}.  With the recent discovery of a Higgs-like
boson \cite{Aad:2012tfa,Chatrchyan:2012ufa} consistent with the
standard model global fits, the need for a higher precision
measurement $W$ boson shifts to physics beyond the standard model.
Models with enhanced symmetries, such as supersymmetry, predict shifts
of 2--20 MeV \cite{Heinemeyer:2006px,Baak:2013fwa}, hence the mass of
the $W$ boson is an important constraint on these models.

Current theoretical predictions of the $W$ boson mass in the standard
model include the full two-loop corrections \cite{Awramik:2003rn} and
leading 3- and 4-loop corrections
\cite{vanderBij:2000cg,Faisst:2003px,Schroder:2005db,Chetyrkin:2006bj,Boughezal:2006xk}.
The current standard model uncertainty is estimated to be $\sim$4~MeV,
however inclusion of full 3-loop self-energies should reduce this to
$\sim$1~MeV \cite{Baak:2013fwa}.  In supersymmetry, the shifts in mass
can be large, but the additional uncertainty due to higher order
effects tends to be small \cite{Heinemeyer:2006px}, hence an
experimental precision of 5~MeV or better is desirable to constrain
the supersymmetric parameter space
\cite{Baak:2013fwa,Beneke:2000hk,MoortgatPick:2005cw}.

The experimental uncertainty $W$ mass measurement from the Tevatron
was well balanced between systematic errors, predominantly lepton
energy scale and recoil energy resolutions, and uncertainties due to
proton structure through the PDFs
\cite{Aaltonen:2013iut,Aaltonen:2013vwa}.  After combining CDF and D0
data sets, the PDF uncertainties remained the largest single
uncertainty at $\pm 10$ MeV on their own.  At the LHC the statistical
errors will be negligible, and the systematics are predicted to be
well under control \cite{Buge:2006dv,Besson:2008zs} --- at the 2--4\%
level that ATLAS found in measuring the $W$ transverse momentum
spectrum \cite{Aad:2011fp}.  However, current experimental estimates
of the uncertainty due to PDFs at the LHC utilized by the ATLAS and
CMS Collaborations are $\pm 25$~MeV \cite{Buge:2006dv}.  Hence,
without improvement in the PDF uncertainties, a measurement at the LHC
will not contribute significantly to the world average.

Recently there have been claims that the errors are severely
overestimated, and are closer to $\pm 10$~MeV at the LHC using current
techniques \cite{Bozzi:2011ww,Rojo:2013nia}.  Further, there are
predictions that the uncertainty will reach $\pm 5$~MeV at the LHC
with expected improvements in the measurement of the PDFs
\cite{Baak:2013fwa}.  We demonstrate below the current uncertainty
predictions are actually slightly underestimated, but we provide a
method to reach $\pm 10$~MeV using current PDF uncertainties and 7 TeV
or 13 TeV data with 90\% confidence level.  As PDF uncertainties
improve, the goal of $\pm 5$~MeV may still be in reach.

We begin our exploration of PDF errors in Sec.\ \ref{sec:methods} by
describing the methods used to determine the $W$ boson mass, and our
simulation of these methods and the PDF uncertainty.  In Sec.\
\ref{sec:sources} we first demonstrate that uncertainties in proton
structure are important at both a hard scattering level and in their
contribution to soft showering.  Hence, error estimates that ignore some
of these effects are too small.  The most recent Tevatron analyses
have taken the sum of these effects into account, and so we 
reproduce the CDF analysis \cite{Aaltonen:2013vwa} as a check on our
calculations.

We use our full analysis in Sec.\ \ref{sec:LHC} to determine the
current PDF error contribution to the $W$ boson mass uncertainty at
the LHC for 7 and 13~TeV.  We find the current uncertainty is at least
$\pm 30$~MeV, far above the desired range, and hence examine a few
strategies in Sec.\ \ref{sec:improve} that can be used to reduce the
PDF contribution to $\pm 10$~MeV or below.  Finally, we conclude in
Sec.\ \ref{sec:concl} with a discussion of where improvements are
needed to reach $\pm 5$~MeV.

\section{Determining the W mass}
\label{sec:methods}

The $W$ boson is an unstable particle, decaying either into jets or a
charged lepton and neutrino. At hadron colliders, observation of $W$
decay into jets is extremely difficult near threshold due to the
resolution of reconstructing jets. The lepton channel ($l = e$, $\mu$)
is much cleaner in this environment. In contrast to the $Z$, one of
the decay products in this channel is invisible: the
neutrino. Therefore, one cannot directly reconstruct the mass of the
decay pair. However, distributions in the observable charged lepton
show kinematic edges sensitive to the $W$ mass. Typically experiments
fit the full shaped of the observed distributions to templates to
determine the best fit mass. Three such observables are commonly used:

\begin{itemize}
\item Transverse momentum ($p_T^l$) of the charged lepton. This is the
  simplest distribution to reconstruct, but suffers from the
  disadvantage that it is sensitive to the underlying $p_T$
  distribution of the $W$ itself. Modeling this distribution requires
  careful attention to nonperturbative, resummed perturbative, and
  fixed-order corrections depending on the $p_T^W$ regime. In
  practice, an experiment will restrict to small $p_T^W$ to avoid the
  poorer resolution at higher $p_T$. This means that, in principle, a
  resummed generator with nonperturbative effects like ResBos
  \cite{Ladinsky:1993zn,Balazs:1997xd,Landry:2002ix} is ideal.  In our
  analysis we are only interested in the relative shift in shape due
  to changes in PDF assumptions.  Therefore, we use a matched
  fixed-order plus parton shower, MadGraph \cite{Alwall:2011uj} plus
  PYTHIA \cite{Sjostrand:2006za}, which is sufficient to describe the
  general features of the $W$ mass analysis, and offers the advantage
  of generating events for analysis in a convenient format. We will
  discuss the details of our modeling in a later section.

\item Missing transverse energy ($E_T^{miss}$ = $p_T^{\nu}$). The
  neutrino transverse momentum is reconstructed by adding all other
  transverse energy in the detector and requiring the total transverse
  momentum to sum to zero, i.e.,
\be
\vec{E}_T^{miss} = -\sum(\vec{E}_T) .
\ee
This observable suffers from the disadvantages of $p_T^l$, and
additionally, any error in the measurement of every other particle in
the detector, hadrons in particular. As such, it is poor variable to
fit, particularly in an active environment such as the LHC, and will
not be considered further.

\item Transverse mass of the $W$. Transverse mass is defined as
\be
M_T = \sqrt{2p_T^l E_T^{miss} (1-\cos(\Delta \phi_{l,miss}))}.
\ee
where $\Delta \phi_{l,miss}$ denotes the angular separation between
the charged lepton and reconstructed neutrino in the transverse plane.
$M_T$ has a Jacobian peak sensitive to the mass of the $W$ without
being dependent on the unobserved longitudinal momentum of the
neutrino, and is much less sensitive to the underlying $W$ $p_T$
distribution than the leptons themselves.

\end{itemize}

Each of these observables has a slightly different shape depending on
the precise mass of the $W$. In particular, in the limit of a
tree-dominated, zero width process with a perfect detector, the
observables would have a steep drop-off at $p_T = M_W/2$ and $M_T =
M_W$. In practice these distributions are much smoother (and easier to
fit) once realistic effects are included.

The $W$ mass is fit by histogramming one of these variables and
comparing to a template, using a best-fit $\chi^2$ or likelihood
function. We use the $\chi^2$ method, using the statistical error in
each bin for a typical number of events as the measure of the fit.

\subsection{Analysis setup}
\label{sec:setup}

Since we are only interested in PDF errors and their origins, we only
need enough realism to reproduce results seen in experimental
analyses; we are not performing an actual mass fit to a template with
full detector effects. To isolate the effect of parton distributions
with as simple an analysis as possible, we do the following: We
generate pseudodata with a similar number of events as existent or
anticipated experimental analyses. In practice this finite data would
be compared to a parametrized template fit from billions of simulated
events, or equivalently to histograms of those billions of events
themselves. We create templates out of the pseudodata, and reweight
event-by-event for different hypothesis masses and PDFs. This forces
the best fit for the central PDF (that used in the original
generation) to occur at the generated mass, which we have chosen to be
$M_W = 80.4$ GeV. 

We can then find PDF errors by comparing the best fit for each PDF
eigenvector to the closest matching mass template.  The total
asymmetric error is found by summing the positive and negative shifts
in mass for each eigenvector in quadrature according to the standard
CTEQ ``modified tolerance method''
\cite{Sullivan:2001ry,Sullivan:2002jt},
\begin{eqnarray}
\delta M_W^{\pm} & = & \sqrt{\sum_{i=1}^{n}\Bigl( \max 
[\, \pm(M_W^i-M_W^0), \pm(M_W^{-i}-M_W^0),0]\Bigr)^2}
\; ,\label{eq:pdferr}
\end{eqnarray}
where $n$ is the number of eigenvectors in the error set, and $M_W^i$ is
the reconstructed mass assuming PDF error set $i$.

In order for the templates to not be sensitive to the statistical
fluctuations of the pseudodata, the entire event must retain the same
fluctuations, from shower history to detector smearing. Therefore we
implement a custom detector simulation that captures most of the
realism while allowing us to maintain control over the random numbers
used to generate the smearing. Otherwise our PDF errors would include
contributions proportional to the errors arising from statistics and
detector systematics.

\subsection{Detector simulation}
\label{sec:detector}

Our ``detector'' consists of a set of calorimeters which smear the
momenta of the particles of the simulated event. Parameters for the
CDF EM calorimeter are taken from Ref.~\cite{Aaltonen:2013vwa}. In
comparing to their analysis (Sec.~\ref{sec:sources}) we model the
reconstructed hadronic recoil according to parameters of that
reference as well, including the min bias contribution. For the
purposes of comparing to their analysis we estimated the PYTHIA
parameter PARP(131) = 0.1 for the luminosity considered there, which
gives approximately 3-4 interactions per crossing. Other details of
the event generation can be found in Sec.~\ref{subsec:events}. Our
generated distributions reproduce those in
Ref.~\cite{Aaltonen:2013vwa} quite well. For the LHC, we use the
parameters of the simulator DELPHES \cite{Ovyn:2009tx} for ATLAS
distributed with MadGraph.

\subsection{Data generation}
\label{subsec:events}

We generate 0--2 jet matched samples using MadGraph 5
\cite{Alwall:2011uj} and shower using PYTHIA 6.420
\cite{Sjostrand:2006za} with a $p_T$-ordered shower. The samples are
matched with the MLM scheme at a scale of 20 GeV. We find that the
matched sample reproduces well the measured $p_T$ distribution of the
$W$ measured by ATLAS \cite{Aad:2011fp}, in contrast to a
pure-showered 0-jet sample, though experiments usually put an upper
cut $p_T^W$ below the matching scale we have used. The matched sample
up to two jets allows every type of parton to participate in $W$
production.  In principle, higher-orders in perturbation theory would
improve the overall normalization, but we find it is less important
once distributions are normalized. Resummation calculations such as
ResBos use fewer arbitrary parameters than PYTHIA tunes in fitting
very low $p_T$, but the $W$ observables under consideration are not
very sensitive to this region, $M_T$ in particular.  The shower does
an excellent job of reproducing the $W$ $p_T$ data at moderate ($> 5$
GeV) $p_T$, as expected, so the generated lepton $p_T$ distribution
should be suitable for the purposes of probing PDF sensitivity.  Data
binned below 5 GeV are unavailable, presumably due to the difficulty
in measuring $W$ recoil from a single lepton plus missing energy in
this region.  However, PYTHIA-showered predictions have been found to
agree with data for the $Z$ recoil down to 2 GeV \cite{Aad:2014xaa}.

As we will see in the next section, the $M_T$ distribution is
sensitive to detector effects through missing energy mismeasurement,
the resolution of which is driven by hadron calorimetry. Since this
effect is so important, a shower, ideally with hadronization, is
needed for a study of $M_T$. ResBos provides a predicted $p_T^W$
distribution and does not resolve individual partons in the
shower. The $p_T^l$ distribution is not sensitive to this issue, only
to the underlying $W$ transverse momentum, and therefore a careful
resummation calculation would serve as a useful check on the shower
evolution for the $p_T^l$ fit. Unfortunately, PDF error eigenvector
grids for ResBos are not available for the LHC and modern PDFs at this
time.

Tevatron samples are generated using the set CTEQ 6.6
\cite{Nadolsky:2008zw}, and LHC samples with CT10
\cite{Lai:2010vv}. PYTHIA has been modified to use these sets via
LHAPDF 5 \cite{Whalley:2005nh}. The PYTHIA tunes used are D6 for the
Tevatron and AMBT1 for the LHC. These tunes are paired with different
PDF sets than those used in their calibration; while this may cause
the low-energy physics to be somewhat different, we are most
interested in the predicted errors for modern PDFs and use them in the
shower for consistent reweighting. 

The $W$ events are decayed to
either $e^- \bar{\nu}_e$/$e^+ \nu_e$ for simplicity; the backgrounds
in the electron channel are negligible in contrast to the muon
channel. Parton uncertainties and strategies for dealing with them
should be similar in the two channels.

\section{Sources of parton distribution error}
\label{sec:sources}

Here we reproduce the latest CDF $W$ mass analysis
\cite{Aaltonen:2013vwa} with simulated Tevatron data to test the rigor
of our simulation. In so doing, we can illuminate the sources of PDF
uncertainty in the $W$ mass fit. We use all detector parameters,
histogram bins, and cuts from that reference. In particular, we follow
the CDF $W$ recoil model in reconstructing the missing transverse
energy. We get a good reproduction of their $W$ recoil spectrum, with
an average $u_T = 5.93 \pm 3.45$ GeV (c.f. $5.92 \pm 3.52$ GeV from
Fig.\ 35 of Ref.\ \cite{Aaltonen:2013vwa}).  After cuts, our sample
contains approximately 440000 events, matching the sample used in the
CDF analysis.

\begin{figure}[!ht]
  \includegraphics[width=3.25in]{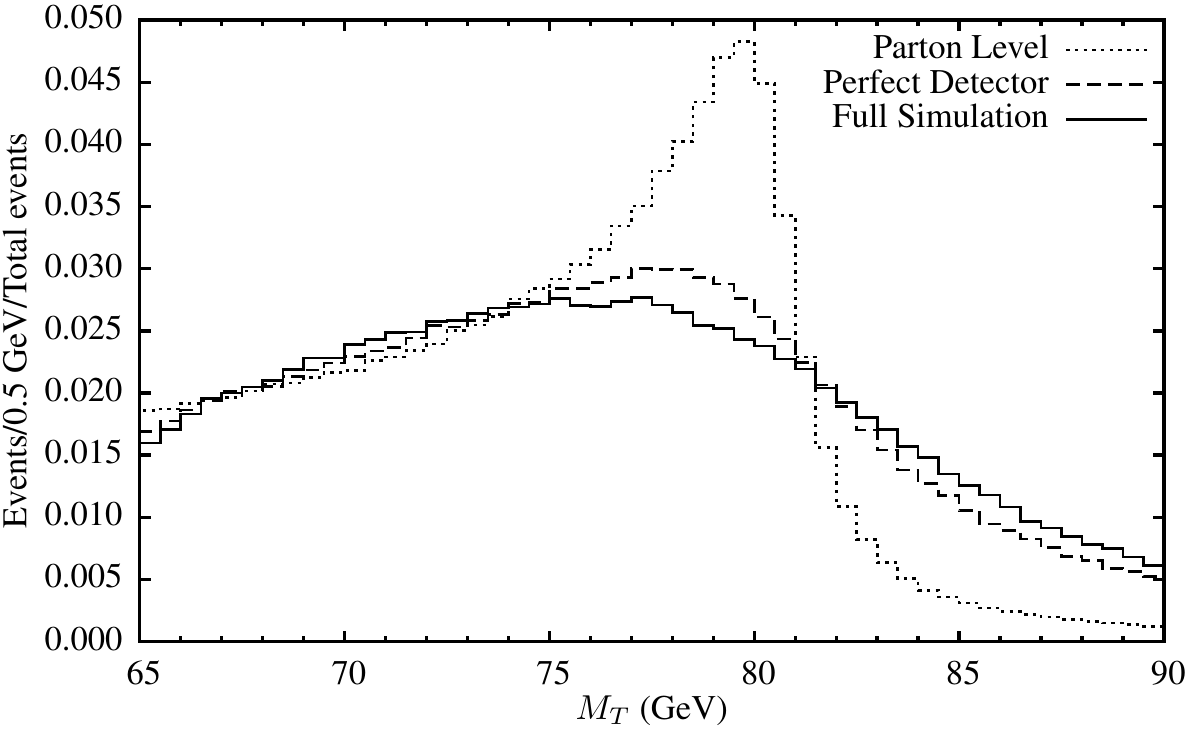}
  \includegraphics[width=3.25in]{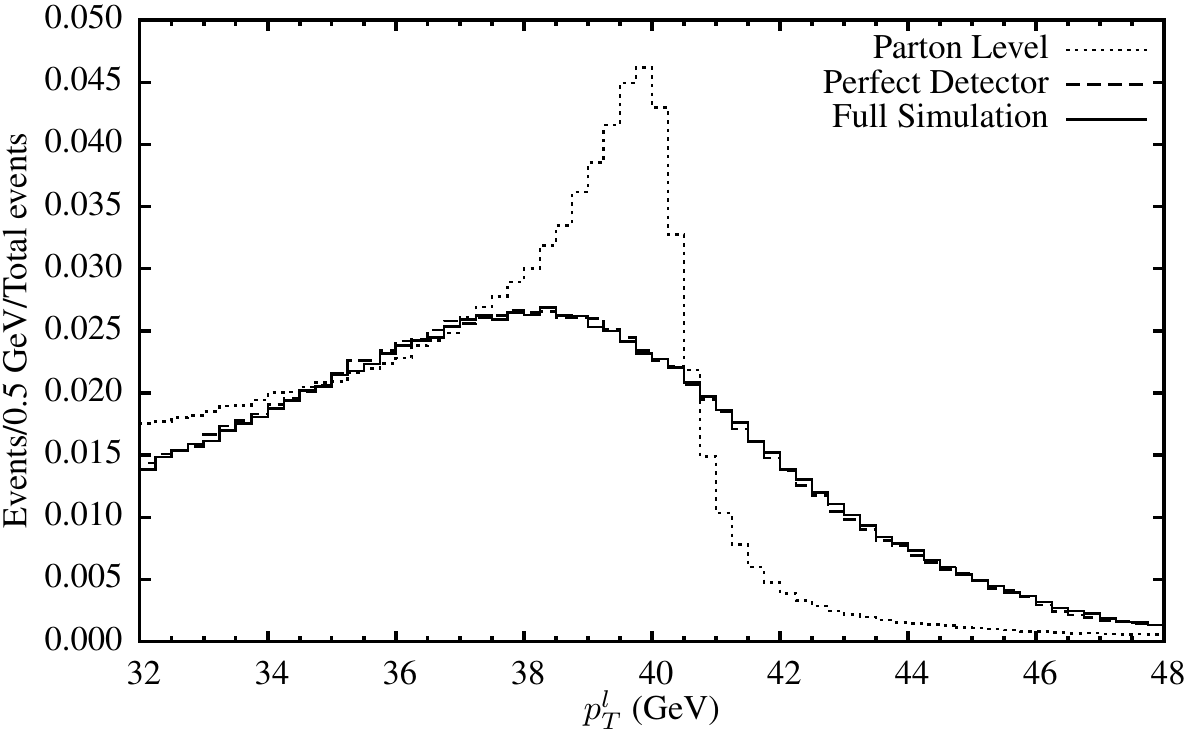}
\caption{Transverse mass and transverse momentum (lepton) distributions
for the Tevatron at three levels of detail in simulation: parton level,
hadron level (perfect detector), and full simulation including detector
and reconstruction efficiencies.
\label{Tev_mtpt} }
\end{figure}

In Fig.~\ref{Tev_mtpt} we plot the transverse mass and transverse
momentum distributions to be fit with increasing layers of realism. We
see that the Jacobian peak present in both distributions at parton
level is badly eroded by the transverse momentum of the $W$, $p_T^e$
especially. $M_T$ also suffers from the missing energy reconstruction
once full detector effects are applied.
Parton distributions affect $M_T$ and $p_T^e$ through acceptance
effects. They alter the rapidity distribution of the produced $W$. The
more central the $W$, the more likely the charged lepton is to decay
in the detector acceptance. Low rapidity $W$ bosons may decay
perpendicular to the beam, allowing for a transverse mass/momentum
near the Jacobian peak. Higher rapidity $W$ bosons must have the
charged lepton decay back toward the detector to be seen, biasing
events away from the Jacobian peak.

We compare the PDF uncertainties of the full simulation to
Ref.~\cite{Aaltonen:2013vwa} using MSTW2008 NLO 68\%CL
\cite{Martin:2009iq} error distributions and find good agreement in
both fits, 10 and 9 MeV, for $M_T$ and $p_T^e$, respectively. From
this point forward we use CTEQ distributions for consistency in
comparison; we are interested in the differences between CT10 and
CT10W \cite{Lai:2010vv} at the LHC in particular.  Note, we make no
attempt to rescale the CTEQ uncertainties in this paper to agree with
the older PDF sets used above, as the PDF fits themselves are not
statistical distributions, and are subject to important systematic
shifts we examine in Sec.\ \ref{sec:LHC}.  Hence, while the numbers
that follow may seem slightly larger, they are an accurate
representation of the current status of the fits.

\begin{table}[!htb]
  \caption{Predicted PDF errors in the Tevatron fit using CTEQ 6.6, with 
increasing levels of realism. All errors in MeV.\label{tab:Tev_err}}
\begin{ruledtabular}
\begin{tabular}{cccc}
& Parton level & + shower & + detector \\ \hline
$M_T$ error & $^{+8}_{-8}$ & $^{+14}_{-13}$ & $^{+18}_{-16}$ \\\hline
$p_T^e$ error & $^{+8}_{-8}$ & $^{+22}_{-20}$ & $^{+22}_{-20}$
\end{tabular}
\end{ruledtabular}
\end{table}

In Table~\ref{tab:Tev_err} we see the effects on the uncertainty due
to the layers of realism, using CTEQ 6.6 PDF uncertainty sets.
There are two main drivers of the increase in PDF error. First, the
smoothing of the distributions to be fit makes it easier to fake a
different mass by shifting PDFs. A perfect Jacobian peak would be
impossible to shift through acceptance effects. Second, PDFs affect
the distribution of $p_T^W$. In fact, especially for the LHC analysis,
we will see that a main contributor of the error in $p_T^e$ comes
through shifting $p_T^W$. To some extent this would be mitigated
through modeling the $W$ recoil by calibrating to the $Z$ recoil, but
it remains to be seen to what extent the PDF shifts in each
distribution correlate.

We have seen that the shower contributes substantially to the PDF
error. This occurs due to the: smoothing of the distributions;
sensitivity to the distribution $p_T^W$; and probing of PDFs at
high-$x$, low $Q^2$ in the shower. Care is taken in our reweighting
procedure to correctly match the scale of the final, lowest $p_T$
emission in the PYTHIA event history. Reweighting at the hard scale,
especially using hard-scale partons, (or equivalently, generating
hard-scale events with different PDF eigenvectors and showering with a
fixed PDF) underestimates the PDF error. The $p_T^W$ distribution is
probing low-scale PDFs.

Some previous theory analyses have missed these effects
by using fixed-order calculations without a detector, and as a result,
dramatically underestimate the resulting errors. It is stated that
only normalization of the distributions is needed to achieve small
residual uncertainties.  We point out that in practice one always
normalizes distributions to the number of events measured in
experimental analyses, cf.\ Ref.\ \cite{Aaltonen:2013vwa}.  It is
true, for the sharply peaked parton-level events, normalization
results in a distribution that is insensitive to changes of the PDFs,
but at the reconstruction level it is not sufficient to reduce
uncertainties at the LHC to a level comparable to the Tevatron.  In
our analyses below we always normalize the distributions and focus on
the sensitivity to shape.

\section{Errors at the LHC}
\label{sec:LHC}

In preparation for the next round of measurements at the LHC we turn to
addressing two questions: what is a realistic estimate of current PDF
uncertainties in a $W$ mass measurement?  And can we achieve the
desired reach of 5 MeV or better?  At the LHC, experimental systematic
uncertainties on the $W$ mass measurement are expected to be under
good control, at around 7 MeV \cite{Besson:2008zs}; statistical errors
are negligible for luminosities in the inverse femtobarn range.  PDF
errors are expected to be the dominant uncertainty, 25 MeV
\cite{Buge:2006dv} for a $M_T$ line shape analysis.  Using
correlations of the $W$ and $Z$ rapidity distributions, there are
claims this can drop to as little as 1 MeV \cite{Besson:2008zs} at the
matrix element level.  In Sec.~\ref{sec:improve} we examine strategies
to reduce the error on reconstructed events to a more realistic 10 MeV
level.

Compared to earlier conditions at the Tevatron, the LHC is a high
pileup environment.  The transverse mass measurement becomes
increasingly difficult as pileup increases, as the missing energy
resolution degrades roughly as the square root of the total hadronic
energy in an event.  While tracking may be able to improve on this,
for the LHC we restrict ourselves to the limit where spectator events
can be removed. Thus the 2011 7 TeV run is preferable to the 8 TeV
run; we also propose that a low-luminosity data sample be acquired if
a fit is to be done using 13 TeV data.

For our simulated $W$ events we impose the selection cuts of
Ref.~\cite{Besson:2008zs}: $p_T^l > 20$ GeV, $p_T^{miss} > 20$ GeV,
recoil $< 30$ GeV. For 7 TeV after cuts we have $1.4\times 10^6$ $W^+$
and $7.15\times 10^5$ $W^-$ events, and for 13 TeV we have $1.1\times
10^6$ $W^+$ and $8\times 10^5$ $W^-$ events, corresponding to an
integrated luminosity of about 2.5 $\textrm{fb}^{-1}$ at 7 TeV, and 0.6
$\textrm{fb}^{-1}$ at 13 TeV, respectively.

In Tab.~\ref{tab:LHC_err} we present the expected errors on the
transverse mass and transverse momentum fits for both data sets for
the CT10 and CT10W PDFs.  These values are larger than the 23--25 MeV
predicted for the $M_T$ fits in Refs.\
\cite{Besson:2008zs,Buge:2006dv} because those predictions used an
older CTEQ 6.1 PDF set \cite{Stump:2003yu}.  With CTEQ 6.1 we find
$\pm 24$ MeV as a baseline uncertainty in complete agreement with
those older predictions.  The larger uncertainties in newer PDFs are
due to relaxation of an artificial restriction on the function form of
the strange quark PDF in the older fits \cite{Nadolsky:2008zw}.
Roughly 30\% of the cross section is directly proportional to the $sc$
initial state.  Hence, this increase is simply an effect of a better
estimate of the $s$ PDF uncertainty.

\begin{table}[!htb]
\caption{PDF errors in the LHC fit, with or without intrinsic charm.
All errors in MeV.
\label{tab:LHC_err}}
\begin{ruledtabular}
  \begin{tabular}{ c | c c c | c c c }
& & 7 TeV & & & 13 TeV &\\
& CT10 & CT10W & CT10+IC & CT10 & CT10W & CT10+IC \\ \hline
$m_T$ error & $^{+39}_{-39}$ & $^{+27}_{-27}$ & $^{+39}_{-40}$ & $^{+30}_{-27}$ & $^{+25}_{-24}$ & $^{+30}_{-31}$ \\ \hline
$p_T^e$ error & $^{+59}_{-54}$ & $^{+46}_{-45}$ & $^{+59}_{-65}$ & $^{+54}_{-52}$ & $^{+48}_{-50}$ & $^{+54}_{-65}$
\end{tabular}
\end{ruledtabular}
\end{table}

In addition, we estimate the resulting contribution to the error if
the input scale charm PDF assumption $c = \bar{c} = 0$ is relaxed by
computing the difference between the intrinsic charm (IC) PDF CTEQ
6.6C2 \cite{Nadolsky:2008zw} and the CTEQ 6.6 central set, and adding
it in quadrature with the other errors, in effect treating it as an
additional eigenvector.  While we do not advocate for the existence of
intrinsic charm, we point out the functional form of the $c$ PDF is not
relaxed in CT10 like it is for $s$.  Hence, we caution that charm
contributions to the uncertainty are not entirely accounted
for. Without the improvements to $W$ mass extraction we propose below,
this could add another $\pm 10$ MeV error at the LHC.

\begin{figure}[!htb]
\includegraphics[width=3.25in]{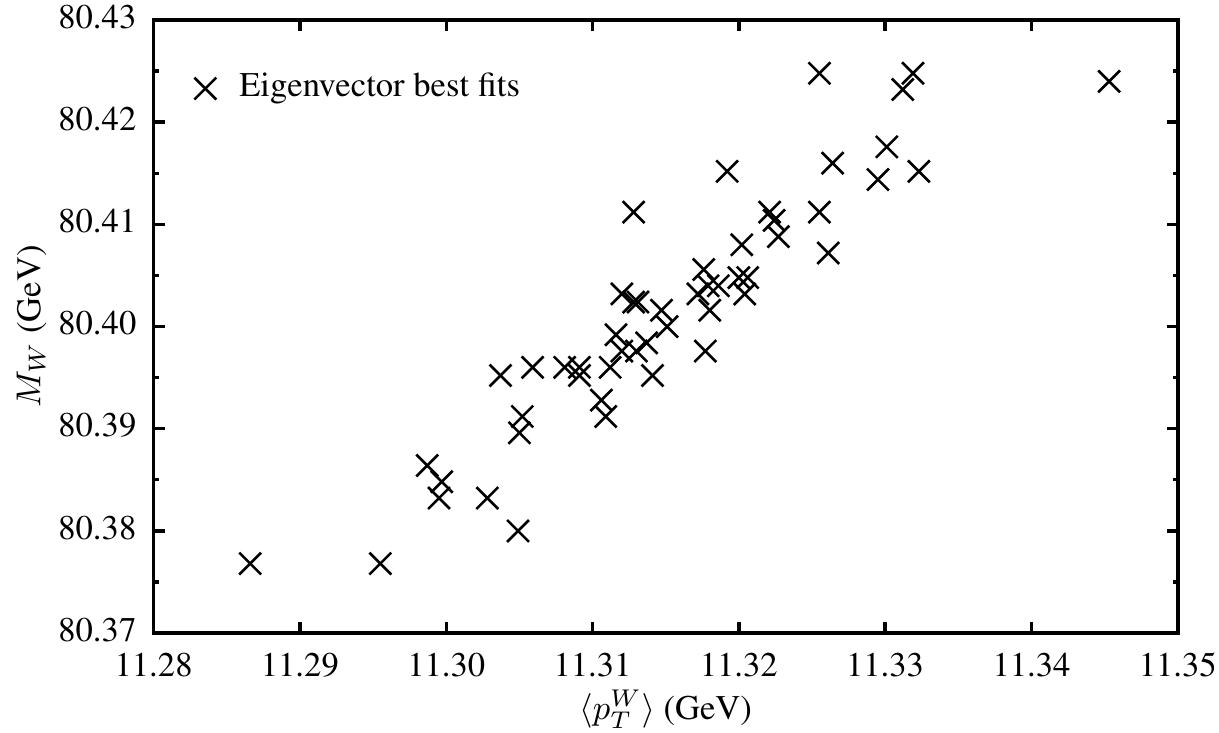}
\caption{The strong correlation between the best fit mass vs.\ average
$W$ transverse momentum.}
\label{fig:ptwcorr}
\end{figure}

We note that expected errors on $p_T^e$ are quite high at the LHC. The
PDFs can induce large shifts in the $W$ recoil, which directly impacts
the $p_T^e$ distribution. To see how this drives the error, we plot
the best-fit $W$ mass versus the average reconstructed $p_T^W$ for
each CT10 eigenvector in Fig.~\ref{fig:ptwcorr}.  The mass shifts are
strongly correlated with the $W$ kick, coming from differences in the
shower and hard emission with PDFs. If the $W$ recoil can be modeled
in some other way (it is usually fit to $Z$ data), this part of the
PDF error can be reduced.

\section{Improving errors}
\label{sec:improve}

Ultimately the parton distributions will be better determined in the
relevant region of $x$ and $Q^2$ using LHC data. Yet the $W$ mass
analysis contains the very same observables that could be used to
constrain those PDFs. We propose to break the analysis into
sub-analyses, each measuring $M_W$. At worst the sub-analysis with the
least sensitivity to PDF uncertainty can be used. As long as there is
not a perfect correlation in the PDF uncertainties among the
sub-analyses, the PDF error will further be improved by combining the
semi-independent analyses. There will still be shared systematic
errors, but the statistical errors are negligible with so many
events.

First, suppose we identify the charged lepton, positron or electron,
and independently measure the mass of $W^+$ and $W^-$. These are of
course equal, but for the purposes of experiment we are measuring two
different quantities with some, but not all, shared errors. What we
are determining is the average
\be
M_W = \frac{M_W^+ + M_W^-}{2} .
\ee
If these are uncorrelated, the error in the average will be better
than the average error, since the errors would be added in
quadrature. At the LHC, $W^+$ and $W^-$ are sensitive to different
PDFs; in addition to sea-produced events, $W^+$ can be produced
through valence $u$ quarks and $W^-$ through valence $d$ quarks at tree
level. These measurements should be at least somewhat uncorrelated.

We can do better by using multiple measurements and taking a weighted
average. Suppose a sub-experiment $i$ measures a $W$ mass
$M_W^i$. Construct an optimized measurement
\be
M_W = \sum_i{\alpha_i M_W^i} \,, \label{eq:submeasure}
\ee
with error
\be
\delta M_W = \sqrt{Var(M_W)} = \sqrt{\sum_i{\alpha_i^2 Var(M_W^i)} + \sum_{i \ne j}{\alpha_i \alpha_j Cov(M_W^i,M_W^j)}} \,,
\ee
where $Var$ and $Cov$ refer to variance and covariance of the errors (PDF
uncertainties in our case) treated as random variables, and $\alpha_i$ are
the weighting coefficients. The overall measurement is subject to the constraint
\be
\sum_i{\alpha_i} = 1 . \label{constraint}
\ee

For sets distributed with independent eigenvectors computed using the
Hessian method, such as CTEQ, one could estimate the covariance using
simple error propagation:
$$
Cov(M_W^i,M_W^j) \approx \sum_k \delta M_W^{i,k} \delta M_W^{j,k} ,
$$
with $\delta M_W^{i,k}$ the deviation of the $i$th measurement due to
eigenvector $k$. However, we choose to use the relation
\be
Var(M_W^i+M_W^j) = Var(M_W^i) + Var(M_W^j) + 2 Cov(M_W^i,M_W^j) ,\label{eq:covar}
\ee
and solve for the covariance, with all variances computed in
accordance with the square of Eq.~\ref{eq:pdferr}: $Var(M_W^i) =
(\delta M_W^i)^2$.  This procedure avoids approximations beyond the
use of the tolerance method, but is complicated by the fact that the
PDF errors are asymmetric. For the purpose of estimating the
covariance matrix we use the average of the $+/-$ error found using
the tolerance method.  In the case of PDF error sets that use a
sampling method, one could compute the variance and covariance
directly using expectation values.

It is straightforward to find the optimum weights to minimize the
error. First, the covariance of two sub-experiments is determined by
adding the shifts of each pair and finding the overall error using
Eq.~\ref{eq:pdferr}. We can then solve for the covariance using
Eq.~\ref{eq:covar}.  If the eigenvector shifts tend to be in opposite
directions for each sub-experiment, they partially cancel, the error
is smaller, and the covariance is negative. If they tend to be in the
same direction, the error adds and the covariance is positive. Once we
know how pairs of sub-experiments correlate, we can minimize the error
$\delta M_W$ over all weights $\alpha$ subject to the constraint in
Eq.~\ref{constraint} by adding a Lagrange multiplier $\lambda$:
\begin{eqnarray}
Var(M_W) &=& \sum_i{\alpha_i^2 Var(M_W^i)} + \sum_{i \ne j}{\alpha_i\alpha_j Cov(M_W^i,M_W^j)} - 2 \lambda (\sum_i \alpha_i - 1) ,\\
\frac{\partial Var(M_W)}{\partial \alpha_j} &=& 2 \alpha_j Var(M_W^j) + 2 \sum_{i \ne j} \alpha_i Cov(M_W^i,M_W^j) - 2\lambda = 0 ,\\
\frac{\partial}{\partial \lambda} (\lambda (\sum_i \alpha_i - 1)) &=& 0 .
\end{eqnarray}

Then, using $Var(M_W^j) = Cov(M_W^j,M_W^j)$ and folding this term with
the $i \ne j$ sum, the optimized $\alpha_j$ obey the system of
equations
\be
\sum_i{\alpha_j Cov(M_W^i, M_W^j)} = \lambda , \label{eq:optimum}
\ee
for all $j$.

Besides separating electron and positron events, there is another way
to get a handle on PDF errors. The pseudorapidity distributions of the
leptons are PDF-dependent. Sea quarks have a bias toward more central
$W$ bosons due the initial state symmetry. Valence quarks tend to
produce more forward events because the high-$x$ peak in their
distribution has to hit a low-$x$ sea quark to produce a
$W$. Furthermore, the $d$ valence distributions peak at lower $x$ than
the $u$ valence.

We therefore propose to split the analysis into low- and
high-pseudorapidity regions. The ATLAS crack at $1.3 < |\eta| < 1.6$
is a good place to split regions for study. We are left with four
sub-analyses to measure the $W$ mass:
\begin{itemize}
\item $W^+$, $|\eta_{e^+}| < 1.3$
\item $W^+$, $|\eta_{e^+}| > 1.6$
\item $W^-$, $|\eta_{e^-}| < 1.3$
\item $W^-$, $|\eta_{e^-}| > 1.6$
\end{itemize}
We refer to the lower pseudorapidity events as central, with a
subscript $c$, and the higher as forward, with a subscript $f$.

\subsection{Transverse mass $W$ subanalysis errors}

After breaking the $W$ mass measurement into four sub-analyses, we
see in Tab.\ \ref{tab:sub_an_err} the PDF errors for each sub-analysis
in the transverse mass fit, $W^+_c$, $W^+_f$, $W^-_c$, and $W^-_f$.
Errors are larger for the forward events compared to central, and
$W^+$ events vs.\ $W^-$. PDFs seem to be under best control for sea
quarks at moderate $x$, though the high/low pseudorapidity difference
is less pronounced at 13 TeV. The $W$ asymmetry data in the CT10W sets
predicts markedly reduced errors, especially for forward events.  The
CT10W PDFs correspond to inclusion of $W$ asymmetry data incompatible
with data samples included in CT10.  Therefore, the spread of CT10
vs.\ CT10W should be taken as an indication of the range of
uncertainty in current PDF estimates.  It is encouraging, that the
CT10W values predict generally smaller errors in the $W$ mass
measurement, and suggest that improvements in $W$ asymmetry data may
have a positive impact on the overall $W$ mass measurement in the
future.

\begin{table}[!htb]
\caption{PDF errors on each sub-analysis, in MeV. Refer to 
Tab.~\protect\ref{tab:LHC_err} for errors on the naive analysis using
all events.
\label{tab:sub_an_err} }
\begin{ruledtabular}
  \begin{tabular}{ c | c c | c c }
& \multicolumn{2}{c |}{7 TeV} & \multicolumn{2}{c}{13 TeV} \\
& CT10 & CT10W & CT10 & CT10W \\ \hline
$W^+_c$ & $^{+46}_{-32}$ & $^{+39}_{-28}$ & $^{+41}_{-30}$ & $^{+36}_{-30}$ \\
$W^+_f$ & $^{+98}_{-102}$ & $^{+68}_{-78}$ & $^{+52}_{-52}$ & $^{+41}_{-42}$ \\
$W^-_c$ & $^{+20}_{-14}$ & $^{+17}_{-13}$ & $^{+29}_{-23}$ & $^{+27}_{-21}$ \\
$W^-_f$ & $^{+49}_{-57}$ & $^{+37}_{-50}$ & $^{+24}_{-35}$ & $^{+19}_{-32}$
  \end{tabular}
\end{ruledtabular}
\end{table}

Already we see that merely restricting to the best-known PDF regions,
by cutting forward and positron events, would improve the error
notably. Each sub-analysis in this case is weakly (anti-)correlated. 
In the case of CT10 at 7 TeV:
$$
Corr(M_W^i, M_W^j) = \left( \begin{array}{cccc}
     1 & 0.218 & 0.255 & -0.44 \\
 0.218 &     1 & 0.0279 & -0.104 \\
 0.255 & 0.0279 &     1 & 0.0439 \\
 -0.44 & -0.104 & 0.0439 &     1 \\
\end{array} \right) ,
$$
for $i = 1$, 2, 3, 4, corresponding to $W^+_c$, $W^+_f$, $W^-_c$,
$W^-_f$. The optimal combination using our method above is
$$
\alpha = \left( \begin{array}{c}
0.144 \\ 0.015 \\ 0.716 \\ 0.125
\end{array} \right) ,
$$
which yields an error on the $W$ mass of +19/-12 MeV, an improvement
of about 60\%. Most of the improvement could be obtained by simply
taking the best sub-measurement, $W^-_c$. For 13 TeV, the optimal
weights are more even, but still dominated by $W^-_c$. Table
\ref{tab:Wimp} summarizes improvements for 7 and 13 TeV for both CT10
and CT10W.

\begin{table}[!htb]
\caption{Resulting error on the $W$ mass after optimal sub-experiment
weighting, in MeV.
\label{tab:Wimp}}
\begin{ruledtabular}
  \begin{tabular}{ c | c c }
& CT10 & CT10W \\ \hline
7 TeV & $^{+19}_{-12}$ & $^{+15}_{-11}$ \\
13 TeV & $^{+20}_{-22}$ & $^{+17}_{-21}$
  \end{tabular}
\end{ruledtabular}
\end{table}

The analysis at 7 TeV appears easier to improve than at 13 TeV; the
optimal pseudorapidity cut may be lower for 13 TeV, but further
pseudorapidity binning did not substantially improve the results,
while degrading the statistics to the point where a coarser template
binning was needed. If a low pileup 13 TeV run can be accomplished
with a few inverse femtobarns of luminosity, a slightly better result
might be obtained.

In contrast to Sec.\ \ref{sec:LHC}, the possibility of underestimated
uncertainty due to, e.g. intrinsic charm, does not greatly alter this
process, since the dominant source of error reduction is choosing a
large weight for the lowest error piece, and the shifts due to charm
are usually no more than a few MeV in this region.

\subsection{Incorporating $Z$ data}

It is expected that incorporating $Z$ measurements will reduce the PDF
error on the $W$ mass measurement, either directly, through refitting
the PDFs, or indirectly, through normalizing $W$ observables to the
$Z$ \cite{Giele:1998uh,Buge:2006dv,Besson:2008zs}.  We estimate the
efficacy of PDF error reduction by using a variant of the latter
method.  Specifically, we want to extend our procedure above adding
additional observables to Eq.~\ref{eq:submeasure}. For instance, if
the $Z$ mass were correlated to the $W$ mass, we might want to measure
$\Delta M_{W,Z} = M_W - M_Z$ instead of measuring the $W$ mass
directly. This would offer experimental advantages as well, since the
systematic errors would likely correlate.

We generate $Z$ samples in the same fashion as our $W$ samples in
Sec.~\ref{sec:setup}, except that we restrict the other lepton, in
this case charged and observable, to the detector acceptance of $-4.9
< \eta < 4.9$, and an invariant mass near the $Z$ mass, $66 \textrm{
  MeV} < M_{ll} < 116 \textrm{ MeV}$. The number of $Z$ events is
chosen to approximate the luminosity of the $W$ events, $3\times 10^5$
for 7 TeV, and $2.2\times 10^5$ for 13 TeV.

\begin{figure}[!ht]
  \includegraphics[width=3.25in]{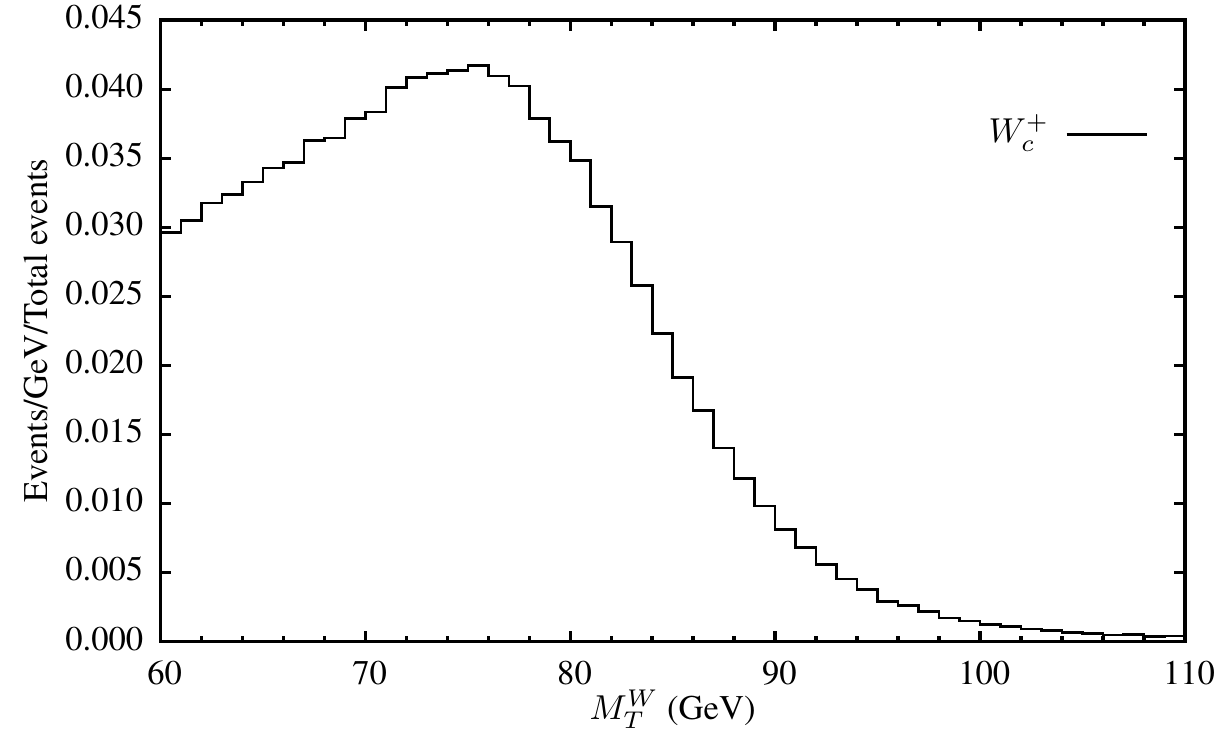}
  \includegraphics[width=3.25in]{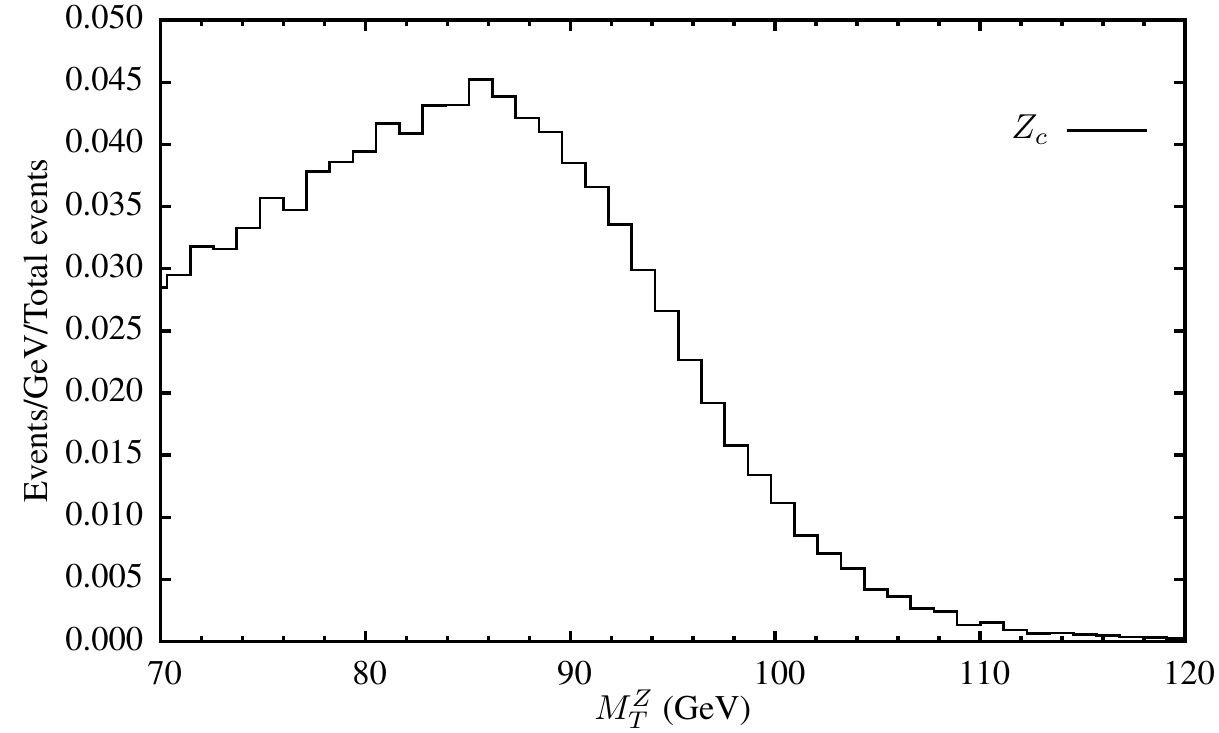}
\caption{Transverse mass distributions for central $W^+$ and $Z$ events at 
the LHC. The histogram windows have been chosen so that the $W$ and $Z$ 
analyses are similar.}
\label{fig:mtwz}
\end{figure}

We follow the same procedure of fitting transverse mass or lepton
transverse momentum from the $Z$ decays to templates.  However, when
we compare the $Z$ fits to the $W$ fits, we rescale the mass or lepton
momentum from the $Z$ measurements by $1/\cos \theta_W = M_Z/M_W$,
whose value should be taken as a prior for the LHC $W$ mass
measurement. Any value will do as long as the histogram windows are
roughly compatible taking the scaling into account.  We plot example
$W$ and $Z$ transverse mass distributions in Fig.~\ref{fig:mtwz}. The
chosen histogram window and binning for the $Z$ is the same as the $W$
but scaled by $1/\cos \theta_W$, hence the similar range on the
$x$-axes. We now have in effect another $W$ mass measurement, $M_W^i =
\cos \theta_W M_Z^i$, for some measurement $M_Z^i$ of the $Z$ mass
done in the same manner as the $W$.

We extend our optimization method using these new measurements $M_W^i$,
which are in fact rescaled measurements of the $Z$ mass. The true
$Z$ mass, whose error is nearly negligible, is used as an input in the
sum $\sum_i{\alpha_i M_W^i}$ for the terms corresponding to $Z$
events. In other words, we are measuring the error on the combination
\be
M_W + \sum_{i \in Z}{\alpha_i M_W^i} = \sum_{i \in W,Z}{\alpha_i M_W^i} , \label{eq:newWmeas}
\ee
where the goal is to minimize the r.h.s. error, and use $M_Z$ as an
input on the l.h.s., whose error is subdominant. The $\alpha_i$ are
unconstrained if $i$ corresponds to a $Z$ measurement, since these
terms appear on both sides of Eq.~\ref{eq:newWmeas}. We can add or
subtract as much or as little of the $Z$ mass as needed to minimize
the combination. Our example $\Delta M_{W,Z}$ above would correspond
to a choice where $\sum_{i \in Z} \alpha_i = -1$. Thus, our
minimization condition Eq.~\ref{eq:optimum} is modified:
\be
\sum_i{\alpha_j Cov(M_W^i, M_W^j)} = 0, j \in Z \label{eq:Zconst};
\ee
the derivative of the lambda term with respect to $\alpha_j$ is zero
if $j \notin W$.

There is no difference between this procedure and the recommended
method of finding the error of the ratio $M_W/M_Z$, since
$$
\left[ \delta\left( \frac{M_W}{M_Z} \right) M_Z\right]^2 \approx (\delta M_W) ^2 
+ \left( \frac{M_W}{M_Z} \delta M_Z\right)^2 - 2 Cov\left(M_W,
 \frac{M_W}{M_Z} M_Z\right) ,
$$
except that our method maintains a simple linear system throughout, and
additional flexibility to modify the contribution of various $Z$
measurements.

Due to the $Z$ vector-axial coupling asymmetry, the positron and
electron can be correlated to different sides of the detector
depending on the overall $Z$ boost, and their distributions would be
sensitive to different PDFs. Therefore we choose to split the $Z$
events into $W^-$-like events, with an electron in the relevant
pseudorapidity region, and $W^+$-like events, with a positron with the
relevant pseudorapidity. The other charged lepton is merely observed
in the detector acceptance for $Z$ identification. We now have four
$Z$ sub-measurements, $Z^+_c$, $Z^+_f$, $Z^-_c$, and $Z^-_f$, one for
each of our $W$ sub-measurements.

It is instructive to examine an older set of PDFs, CTEQ 6.1 first, on
which the studies Refs.\cite{Besson:2008zs,Buge:2006dv} are based. It
is speculated there that $Z$ measurements can nearly eliminate PDF
error due to the especially strong correlations of a parameter
describing the $W$ and $Z$ rapidity distributions; those rapidity
distributions affect the $W$ observables through acceptance effects as
described above.

With CTEQ 6.1 at 13 TeV, we find an especially strong correlation
between certain $W$ and $Z$ sub-measurements:
$$
Corr(M_W^i, M_W^j) = \left( \begin{array}{cccccccc}
    1 & 0.468 & 0.884 & 0.151 & \mathbf{0.849} & -0.29 & \mathbf{0.848} & 0.591 \\
0.468 &     1 &  0.36 & 0.249 & 0.501  &0.186 & 0.375 & 0.642 \\
0.884 &  0.36 &     1 & 0.237 & \mathbf{0.934} &-0.545 &  \mathbf{0.88} & 0.681 \\
0.151 & 0.249 & 0.237 &     1 & 0.064 & 0.343 & 0.097 & 0.286 \\
0.849 & 0.501 & 0.934 & 0.064 &     1 &-0.626 & 0.866 & 0.805 \\
-0.29 & 0.186 &-0.545 & 0.343 &-0.626 &     1 &-0.637 &-0.369 \\
0.848 & 0.375 &  0.88 & 0.097 & 0.866 &-0.637 &     1 & 0.627 \\
0.591 & 0.642 & 0.681 & 0.286 & 0.805 &-0.369 & 0.627 &     1 \\
\end{array} \right) ,
$$
where now the sub-measurements are extended, with the ordering
corresponding to $W^+_c$, $W^+_f$, $W^-_c$, $W^-_f$, $Z^+_c$, $Z^+_f$,
$Z^-_c$, $Z^-_f$. The correlations in bold correspond to the central
region where strong correlations would be expected, since events are
dominated by sea/gluon-initiated processes at the parton level. These
distributions have a strongly constrained functional form, where all
sea distributions are set equal (or zero) at the input scale and
evolve from the gluon distributions beyond mass threshold.

The solution to the system Eqs.~\ref{eq:optimum},~\ref{eq:Zconst} in
this case is
$$
\alpha = \left( \begin{array}{c}
 0.646 \\
 0.197 \\
0.067 \\
0.090 \\
-0.705 \\
-0.391 \\
-0.506 \\
0.026 \\
\end{array} \right) .
$$
As expected, adding $Z$ measurements with negative weight is
optimal. The resulting error is $^{+6}_{-7}$ MeV, a factor of four
improvement! As high as they are, the PDF correlations are not perfect
in $M_T$, so the effect is not as strong as anticipated, however much
the ``spread'' in boson rapidity might correlate.

Even this result is overly optimistic, however. Errors in more modern
sets are much larger, partly a result of the relaxing of functional
constraints such as strangeness starting in CTEQ 6.6. In fact, it is
these very constraints that are causing an overestimate in the $W/Z$
correlation in this and previous analyses.

For comparison, we present the correlations in CT10 at 13 TeV:
$$
Corr(M_W^i, M_W^j) = \left( \begin{array}{cccccccc}
     1 & 0.474 & 0.476 & 0.074 & \mathbf{0.876} &  0.27 & \mathbf{0.412} &-0.153 \\
 0.474 &     1 & 0.062 &-0.018 & 0.552 & 0.521 & 0.107 &-0.069 \\
 0.476 & 0.062 &     1 &  0.33 & \mathbf{0.501} &-0.426 & \mathbf{ 0.730} & 0.458 \\
0.074 &-0.018 &  0.33 &     1 &-0.073 &-0.034 & 0.108 & 0.292 \\
 0.876 & 0.552 & 0.501 &-0.073 &     1 & 0.229 &  0.41 &-0.027 \\
  0.27 & 0.521 &-0.426 &-0.034 & 0.229 &     1 &-0.345 &-0.394 \\
 0.412 & 0.107 &  0.730 & 0.108 &  0.41 &-0.345 &     1 & 0.368 \\
-0.153 &-0.069 & 0.458 & 0.292 &-0.027 &-0.394 & 0.368 &     1 \\
\end{array} \right) .
$$
The relevant correlations have dropped as low as 41\%, compared to 85\% and 
above for CTEQ 6.1. Now, the optimum solution is
$$
\alpha = \left( \begin{array}{c}
 0.401 \\
 0.117 \\
0.560 \\
-0.078 \\
-0.648 \\
0.014 \\
-0.503 \\
0.018 \\
\end{array} \right) .
$$
To good approximation, this is simply taking the central events for
$W$ and $Z$ and subtracting the resulting mass measurements, which was
anticipated above. Due to the imperfect correlations, the error is
reduced only to $^{+10}_{-11}$ MeV; a factor of three improvement over
the larger CT10 error, but not nearly the improvement hoped for.

In Tab.~\ref{tab:errZ} we summarize the end result of our optimization
procedure including $Z$ fits.  It appears that by using the Tevatron
$W$ asymmetry data, and finding the best combination of
sub-experiments possible, including anticorrelations with the $Z$, the
LHC can do no better than 8 MeV without constraining PDFs with other
processes or colliders; this also assumes the current functional form
of CTEQ (and most other distributions) is sufficiently general that no
artificial correlations remain.

\begin{table}[!ht]
\caption{Resulting error on the $W$ mass after optimal sub-experiment
weighting, including $Z$ measurements, in MeV.
\label{tab:errZ} }
\begin{ruledtabular}
  \begin{tabular}{ c | c c }
& CT10 & CT10W \\ \hline
7 TeV & $^{+11}_{-10}$ & $^{+8}_{-8}$ \\
13 TeV & $^{+10}_{-11}$ & $^{+7}_{-11}$
  \end{tabular}
\end{ruledtabular}
\end{table}

To test this hypothesis, we again mimic the uncertainty due to
constraints on the functional form of charm by adding intrinsic charm
as in Sec.~\ref{sec:LHC} to CT10 at 13 TeV, and check the
sub-measurement correlation matrix:
$$
Corr(M_W^i, M_W^j) = \left( \begin{array}{cccccccc}
     1 & 0.507 & 0.492 & 0.108 & \mathbf{0.889} & 0.127 & \mathbf{0.158} &0.021 \\
 0.507 &     1 &0.089 &0.012 & 0.578 & 0.427 &-0.0056 &0.0261 \\
 0.492 &0.089 &     1 & 0.341 & \mathbf{0.518} &-0.443 & \mathbf{0.618} & 0.473 \\
 0.108 &0.012 & 0.341 &     1 &-0.0346 &-0.075 &0.036 & 0.307 \\
 0.889 & 0.578 & 0.518 &-0.035 &     1 & 0.121 &  0.21 & 0.127 \\
 0.127 & 0.427 &-0.443 &-0.075 & 0.121 &     1 & -0.18 &-0.453 \\
 0.158 &-0.005 & 0.618 &0.036 &  0.21 & -0.18 &     1 & 0.142 \\
0.021 &0.026 & 0.473 & 0.307 & 0.127 &-0.453 & 0.142 &     1\\
\end{array} \right) .
$$
The correlations between $W$ and $Z$ have grown weaker still, and the solution
$$
\alpha = \left( \begin{array}{c}
 0.302 \\
 0.104 \\
0.692 \\
-0.098 \\
-0.666 \\
0.093 \\
-0.405 \\
0.060 \\
\end{array} \right)
$$
has shifted slightly. The error is now $^{+10}_{-12}$ MeV, a slightly
worse result. The charm can be compensated for to some extent, but
only if we allow for its existence. Using the same sub-measurement
combination for the original CT10 error set would increase the error
to 14 MeV. Using the $Z$ to cancel the $W$ errors is a bit of a
balancing act and depends on the degrees of freedom present in the
PDFs to begin with. Table~\ref{tab:charmadd} shows the results of
allowing for this additional degree of freedom in the first two
columns, and the result of trying to minimize the error taking into
account the new correlation in the last two columns.

\begin{table}[!htb]
\caption{Resulting error on the $W$ mass including an additional PDF degree
of freedom in charm, with and without optimization for the new degree of
freedom. The ``opt'' column corresponds to the result after re-optimization
of the weights taking the new degree of freedom into account. Since we treat 
intrinsic charm as a single shift without a $+/-$ eigenvector pair, the 
increase shows up asymmetrically.
\label{tab:charmadd} }
\begin{ruledtabular}
  \begin{tabular}{ c | c c | c c }
& CT10 & CT10W & CT10, opt & CT10W, opt\\ \hline
7 TeV & $^{+11}_{-20}$ & $^{+8}_{-24}$ & $^{+13}_{-12}$ & $^{+9}_{-11}$\\
13 TeV & $^{+10}_{-14}$ & $^{+7}_{-12}$ & $^{+10}_{-12}$ & $^{+7}_{-11}$
  \end{tabular}
\end{ruledtabular}
\end{table}

Because of the bigger reliance on sea PDFs for the cancellation at 7
TeV, adding an additional degree of freedom seems to reduce the
effectiveness of the error reduction there significantly.  The large
($\pm 10$ MeV) additional uncertainty is a concern because it
represents a new systematic uncertainty not currently accounted in
other analyses.  Fortunately, the effect almost disappears at 13 TeV,
which is sensitive to a different $x$ and $Q^2$ range, and evidenced
by the small difference between any of the 13 TeV results in Tabs.\
\ref{tab:errZ} and \ref{tab:charmadd}.  Hence, this suggests there is
more control over PDF uncertainties at the higher LHC energy.

\subsection{Lepton transverse momentum}

In Tab.~\ref{tab:LHC_err} we saw that PDF errors are expected to be
considerably larger for lepton transverse momentum than transverse
mass at the LHC, and Fig.~\ref{fig:ptwcorr} shows that this is due to
the variation in $W$ recoil spectrum.
If this is the case, perhaps once again the $Z$ can come to the
rescue. It is usually assumed that the low-energy physics governing
the $p_T$ spectrum of the bosons is universal up to scale; therefore
one should be able to fit the $W$ recoil spectrum, which requires the
badly-measured missing energy, to that of the $Z$, where both
well-measured leptons in the decay can be added to produce $p_T^Z$.

\begin{figure}[!htb]
\includegraphics[width=3.25in]{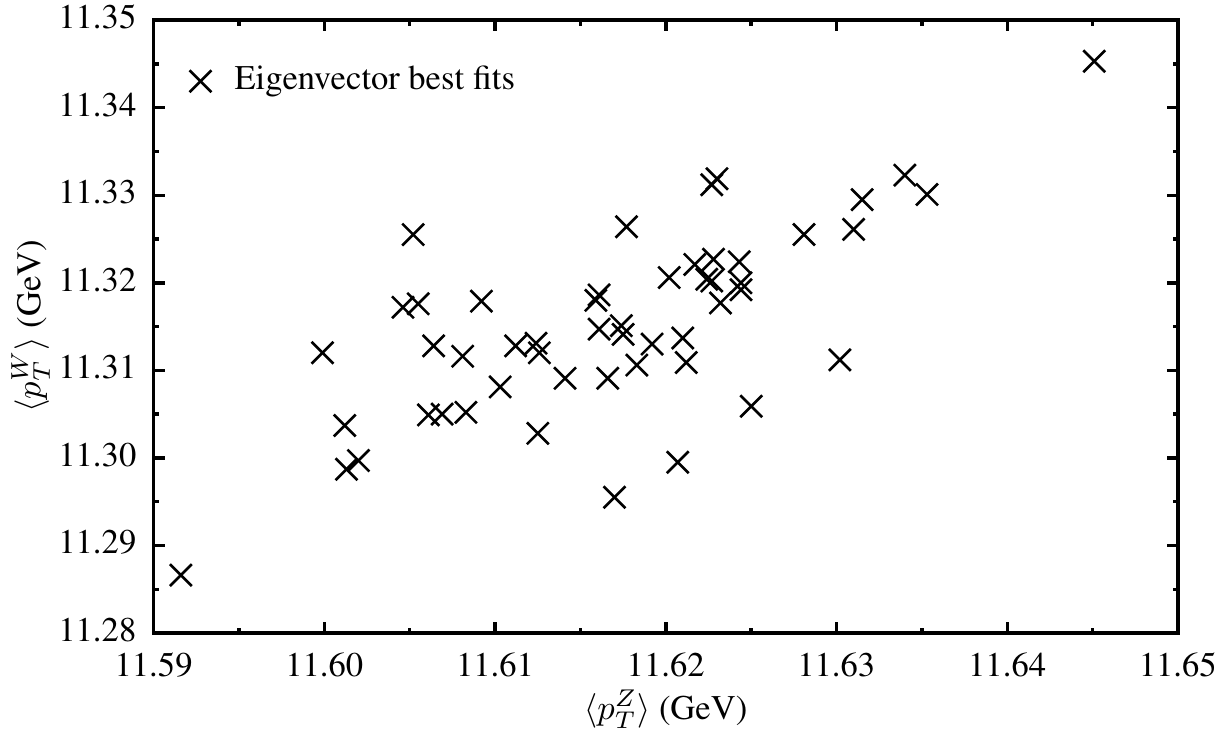}
\caption{Shifts in the $W$ and $Z$ recoil due to each PDF eigenvector.}
\label{fig:ptwzcorr}
\end{figure}

We examine this assumption by plotting the joint $p_T$ shifts of
the bosons due to each PDF eigenvector in Fig.~\ref{fig:ptwzcorr}.
The PDF errors on the mass measurement may be strongly correlated with
the recoil, but the $W$ and $Z$ recoil are certainly not completely
correlated with each other. Fitting one to the other is an assumption
that should be critically reevaluated in the context of PDFs. We use
the predicted correlations in our procedure as before to reduce the
PDF error. Since the $W$ and $Z$ recoil are correlated, the $p_T$ mass
fits should also be, and we can find the optimum combination as in the
transverse mass case.

Below in Tab.~\ref{tab:pt} we present the results of our procedure for
the $p_T^e$ fit. There are generally stronger correlations between $W$
and $Z$ observables than in the transverse mass case, but not enough
to overcome the larger inherent error. $M_T$ appears to be the
preferred variable at the LHC unless pileup becomes a problem.

\begin{table}[!ht]
\caption{Resulting error on the W mass for $p_T^e$, in MeV, after optimizing
with $W$ and $Z$ data.
\label{tab:pt}}
\begin{ruledtabular}
  \begin{tabular}{c | c c }
& CT10 & CT10W \\ \hline
7 TeV & $^{+17}_{-17}$ & $^{+11}_{-14}$ \\
13 TeV & $^{+18}_{-16}$ & $^{+14}_{-15}$
  \end{tabular}
\end{ruledtabular}
\end{table}

\section{Conclusions}
\label{sec:concl}

We have critically reevaluated the contribution of PDFs to the error
on a potential mass measurement of the $W$ boson at the
LHC. Over-optimistic analyses have been shown to rely on unrealistic
distributions in the observables used to constrain the $W$ mass
through template fitting. The softening of distributions through
emission of additional partons and detector mismeasurement result in
deviations that are much easier to reproduce by shifts in PDFs.

We have also devised strategies to combine sub-measurements to reduce
the impact of PDF uncertainties, with and without the addition of $Z$
boson observables. Since shifts in $W$ and $Z$ distributions are
correlated in PDF space, this correlation can be exploited to reduce
the error by also using the more directly measured $Z$ mass as an
input, a generalization of normalizing to the $Z$. However, past
estimates of this effect have been greatly exaggerated due in part to
artificial correlations induced by the assumed shared structure of sea
partons in PDF fits. Adding freedom to the strange quark in more
modern sets not only increases the expected PDF error directly, but
reduces the correlations between $W$ and $Z$ due to the different
parton flavors producing them at the high scale. New degrees of
freedom such as looser restrictions on the fit of charm would increase
the error further, but there do not appear to be strong artificial
correlations remaining after strangeness is added.

In particular, we have shown that the transverse momentum
distributions of $W$ and $Z$ do not entirely correlate in PDF
space. If $Z$ data is used to assume the underlying distribution of
the $W$, care should be taken to propagate the additional PDF errors
correctly; there is a PDF error component to the model of the $W$
recoil. PDF errors in the $p_T^e$ mass fit are large at the LHC due to
PDF effects on the recoil. Transverse momentum should be used for mass
fits to minimize PDF error unless missing energy resolution degrades
too badly.

Having identified both the causes and a solution to the PDF portion of
the $W$ mass uncertainty, the next step should be to integrate these
cuts into a fully resummed calculation including soft photon radiation
effects.  A promising framework has recently been proposed called
DYRES, which adds resummation to the calculations of $W$ and $Z$
production \cite{Catani:2009sm}.  As the performance of the LHC
detectors is understood in Run II, we encourage an full analysis by
the experimental collaborations using the latest tools available.

Our most optimistic estimate of the PDF error contribution on the $W$
mass measurement at the LHC is 8 MeV, using CT10W PDFs, if our method of
utilizing the correlation with $Z$ mass fits is followed.  It is
encouraging that forward (large-$x$) $W$ data can both improve and
stabilize the uncertainties.  However, we recommend a more
conservative $\pm 10$--$12$ MeV as a fair estimate of what can be
currently achieved.  To reach the ultimate goal of $\pm 5$ MeV,
further improvement will require additional data and PDF fits beyond
$W$ and $Z$ production.

\bigskip \bigskip

\begin{acknowledgments}
  This work is supported by the U.S.\ Department of Energy under
  Contract Nos.\ DE-FG02-13ER41942 and DE-SC0008347.
\end{acknowledgments}


\begin{thebibliography}{99}

\bibitem{Baak:2013fwa} 
M.~Baak
{\it et al.},
in  {\sl Planning the Future of U.S.\ Particle
Physics: Report of the 2013 Community Summer Study of the APS Division
of Particles and Fields}, 
edited by Norman A.\ Graf, Michael E.\ Peskin, and Jonathan L.\ Rosner,
eConf C1307292, EF19 (2014)
[arXiv:1310.6708 [hep-ph]].

\bibitem{Aaltonen:2013iut} 
T.~A.~Aaltonen {\it et al.}  [CDF and D0 Collaborations],
Phys.\ Rev.\ D {\bf 88}, no. 5, 052018 (2013)
[arXiv:1307.7627 [hep-ex]].

\bibitem{Arnison:1983rp} 
G.~Arnison {\it et al.}  [UA1 Collaboration],
Phys.\ Lett.\ B {\bf 122}, 103 (1983).



\bibitem{Agashe:2014kda} 
  K.~A.~Olive {\it et al.}  [Particle Data Group Collaboration],
  Chin.\ Phys.\ C {\bf 38}, 090001 (2014).

\bibitem{Aad:2012tfa} 
  G.~Aad {\it et al.}  [ATLAS Collaboration],
  Phys.\ Lett.\ B {\bf 716}, 1 (2012)
  [arXiv:1207.7214 [hep-ex]].

\bibitem{Chatrchyan:2012ufa} 
  S.~Chatrchyan {\it et al.}  [CMS Collaboration],
  Phys.\ Lett.\ B {\bf 716}, 30 (2012)
  [arXiv:1207.7235 [hep-ex]].

\bibitem{Heinemeyer:2006px} 
  S.~Heinemeyer, W.~Hollik, D.~Stockinger, A.~M.~Weber and G.~Weiglein,
  J.\ High Energy Phys.\  {\bf 0608}, 052 (2006)
  [hep-ph/0604147].

\bibitem{Awramik:2003rn} 
M.~Awramik, M.~Czakon, A.~Freitas, and G.~Weiglein,
Phys.\ Rev.\ D {\bf 69}, 053006 (2004)
[hep-ph/0311148].

\bibitem{vanderBij:2000cg} 
J.J.~van der Bij, K.G.~Chetyrkin, M.~Faisst, G.~Jikia, and T.~Seidensticker,
Phys.\ Lett.\ B {\bf 498}, 156 (2001)
[hep-ph/0011373].

\bibitem{Faisst:2003px}
M.~Faisst, J.H.~Kuhn, T.~Seidensticker, and O.~Veretin,
Nucl.\ Phys.\ B {\bf 665}, 649 (2003)
[hep-ph/0302275].

\bibitem{Schroder:2005db} 
Y.~Schroder and M.~Steinhauser,
Phys.\ Lett.\ B {\bf 622}, 124 (2005)
[hep-ph/0504055].

\bibitem{Chetyrkin:2006bj} 
K.G.~Chetyrkin, M.~Faisst, J.H.~Kuhn, P.~Maierhofer, and C.~Sturm,
Phys.\ Rev.\ Lett.\  {\bf 97}, 102003 (2006)
[hep-ph/0605201].

\bibitem{Boughezal:2006xk} 
R.~Boughezal and M.~Czakon,
Nucl.\ Phys.\ B {\bf 755}, 221 (2006)
[hep-ph/0606232].

\bibitem{Beneke:2000hk} 
  M.~Beneke, I.~Efthymiopoulos, M.~L.~Mangano, J.~Womersley, A.~Ahmadov, G.~Azuelos, U.~Baur and A.~Belyaev {\it et al.},
  in {\it Proceedings of the Workshop on Standard Model Physics 818
(and More) at the LHC}, edited by G.\ Altarelli and M.~L.\ 
Mangano (CERN, Geneva, 2000), p.\ 419
  [hep-ph/0003033].

\bibitem{MoortgatPick:2005cw} 
  G.~Moortgat-Pick, T.~Abe, G.~Alexander, B.~Ananthanarayan, A.~A.~Babich, V.~Bharadwaj, D.~Barber and A.~Bartl {\it et al.},
  Phys.\ Rept.\  {\bf 460}, 131 (2008)
  [hep-ph/0507011].

\bibitem{Aaltonen:2013vwa} 
  T.~A.~Aaltonen {\it et al.}  [CDF Collaboration],
  Phys.\ Rev.\ D {\bf 89}, no. 7, 072003 (2014)
  [arXiv:1311.0894 [hep-ex]].

\bibitem{Buge:2006dv} 
  V.~Buge, C.~Jung, G.~Quast, A.~Ghezzi, M.~Malberti and T.~Tabarelli de Fatis,
  J.\ Phys.\ G {\bf 34}, N193 (2007).


\bibitem{Besson:2008zs} 
  N.~Besson {\it et al.}  [ATLAS Collaboration],
  Eur.\ Phys.\ J.\ C {\bf 57}, 627 (2008)
  [arXiv:0805.2093 [hep-ex]].

\bibitem{Aad:2011fp}
  G.~Aad {\it et al.}  [ATLAS Collaboration],
  Phys.\ Rev.\ D {\bf 85}, 012005 (2012)
  [arXiv:1108.6308 [hep-ex]].

\bibitem{Rojo:2013nia} 
J.~Rojo and A.~Vicini,
arXiv:1309.1311 [hep-ph].

\bibitem{Bozzi:2011ww} 
  G.~Bozzi, J.~Rojo and A.~Vicini,
  Phys.\ Rev.\ D {\bf 83}, 113008 (2011)
  [arXiv:1104.2056 [hep-ph]].




\bibitem{Ladinsky:1993zn}
  G.~A.~Ladinsky and C.~P.~Yuan,
  Phys.\ Rev.\ D {\bf 50}, 4239 (1994)
  [hep-ph/9311341].

\bibitem{Balazs:1997xd}
  C.~Balazs and C.~P.~Yuan,
  Phys.\ Rev.\ D {\bf 56}, 5558 (1997)
  [hep-ph/9704258].

\bibitem{Landry:2002ix}
  F.~Landry, R.~Brock, P.~M.~Nadolsky and C.~P.~Yuan,
  Phys.\ Rev.\ D {\bf 67}, 073016 (2003)
  [hep-ph/0212159].

\bibitem{Alwall:2011uj} 
  J.~Alwall, M.~Herquet, F.~Maltoni, O.~Mattelaer and T.~Stelzer,
  J.\ High Energy Phys.\  {\bf 1106}, 128 (2011)
  [arXiv:1106.0522 [hep-ph]].

\bibitem{Sjostrand:2006za} 
  T.~Sjostrand, S.~Mrenna and P.~Z.~Skands,
  J.\ High Energy Phys.\  {\bf 0605}, 026 (2006)
  [hep-ph/0603175].

\bibitem{Sullivan:2001ry}
Z.~Sullivan and P.M.~Nadolsky, ``Heavy-quark parton distribution
functions and their uncertainties'' in {\sl Proceedings of Snowmass 2001: the
Future of Particle Physics}, Snowmass, July 1--20, 2001, edited by N.~Graf
(SLAC, Stanford, 2002), eConf C010630, P511
[hep-ph/0111358].

\bibitem{Sullivan:2002jt} 
  Zack~Sullivan,
  Phys.\ Rev.\ D {\bf 66}, 075011 (2002)
  [hep-ph/0207290].

\bibitem{Ovyn:2009tx}
  S.~Ovyn, X.~Rouby and V.~Lemaitre,
  arXiv:0903.2225 [hep-ph].

\bibitem{Aad:2014xaa} 
  G.~Aad {\it et al.} [ATLAS Collaboration],
  J.\ High Energy Phys.\ {\bf 1409}, 145 (2014)
  [arXiv:1406.3660 [hep-ex]].



\bibitem{Nadolsky:2008zw} 
  P.~M.~Nadolsky, H.~L.~Lai, Q.~H.~Cao, J.~Huston, J.~Pumplin, D.~Stump, W.~K.~Tung and C.-P.~Yuan,
  Phys.\ Rev.\ D {\bf 78}, 013004 (2008)
  [arXiv:0802.0007 [hep-ph]].

\bibitem{Lai:2010vv}
  H.~L.~Lai, M.~Guzzi, J.~Huston, Z.~Li, P.~M.~Nadolsky, J.~Pumplin and
C.-P.~Yuan,
  Phys.\ Rev.\ D {\bf 82}, 074024 (2010)
  [arXiv:1007.2241 [hep-ph]].

\bibitem{Whalley:2005nh}
  M.~R.~Whalley, D.~Bourilkov and R.~C.~Group,
  hep-ph/0508110.



\bibitem{Martin:2009iq}
  A.~D.~Martin, W.~J.~Stirling, R.~S.~Thorne and G.~Watt,
  Eur.\ Phys.\ J.\ C {\bf 63}, 189 (2009)
  [arXiv:0901.0002 [hep-ph]].




\bibitem{Stump:2003yu} 
  D.~Stump, J.~Huston, J.~Pumplin, W.~K.~Tung, H.~L.~Lai, S.~Kuhlmann and J.~F.~Owens,
  J.\ High Energy Phys.\ {\bf 0310}, 046 (2003)
  [hep-ph/0303013].

\bibitem{Giele:1998uh} 
  W.~T.~Giele and S.~Keller,
  Phys.\ Rev.\ D {\bf 57}, 4433 (1998)
  [hep-ph/9704419].

\bibitem{Catani:2009sm} 
  S.~Catani, L.~Cieri, G.~Ferrera, D.~de Florian and M.~Grazzini,
  Phys.\ Rev.\ Lett.\  {\bf 103}, 082001 (2009)
  [arXiv:0903.2120 [hep-ph]].




\end{thebibliography}
\end{document}